\documentclass[lettersize,journal]{IEEEtran}
\usepackage{tikz}
\usepackage[utf8]{inputenc}
\usepackage{graphicx}
\usepackage[caption=false]{subfig}
\usepackage{caption}
\usepackage{verbatim}
\usepackage{makecell}
\usepackage{cite}
\usepackage{multirow}
\usepackage{circuitikz}
\usepackage{mathtools}
\usepackage{nccmath}
\definecolor{accessblue}{RGB}{0,105,154}
\usepackage{amsmath,amssymb,amsfonts}
\usepackage{algorithmic}
\usepackage{textcomp}
\usepackage{longtable}
\usepackage{array}
\usepackage{etoolbox}
\usepackage[normalem]{ulem}
\usepackage{xcolor}
\usepackage{lipsum}
\begin{document}

\title{Power Flow Analysis Using Deep Neural Networks in Three-Phase Unbalanced Smart Distribution Grids}

\author{{Deepak Tiwari, Mehdi~Jabbari Zideh, Veeru Talreja, Vishal Verma, Sarika K. Solanki, and Jignesh Solanki}


\thanks{D. Tiwari is with Commonwealth Edison (ComEd), Chicago, USA (e-mail: Deepak.Tiwari@ComEd.com)}

\thanks{M. Jabbari Zideh, V. Talreja, S. K. Solanki, and J. Solanki are with the Lane Department of Computer Science and Electrical Engineering, West Virginia University, Morgantown, WV 26505 USA (e-mail: mj00021@mix.wvu.edu, vtalreja@mix.wvu.edu, skhushalanisolanki@mail.wvu.edu, jignesh.solanki@mail.wvu.edu)}

\thanks{V. Verma is with Electric Power Research Institute, USA (e-mail: vishal.0962@gmail.com)}


\thanks{\emph{Corresponding author: Mehdi Jabbari Zideh.}}}
\maketitle
\begin{abstract}   
Most power systems' approaches are currently tending towards stochastic and probabilistic methods due to the high variability of renewable sources and the stochastic nature of loads. Conventional power flow (PF) approaches such as forward-backward sweep (FBS) and Newton-Raphson require a high number of iterations to solve non-linear PF equations making them computationally very intensive. PF is the most important study performed by utility, required in all stages of the power system, especially in operations and planning. This paper discusses the applications of deep learning (DL) to predict PF solutions for three-phase unbalanced power distribution grids. Three deep neural networks (DNNs); Radial Basis Function Network (RBFnet), Multi-Layer Perceptron (MLP), and Convolutional Neural Network (CNN), are proposed in this paper to predict PF solutions. The strength of the proposed DNN models over the traditional iterative-based PF solvers is that these models can capture the nonlinear relationships in PF calculations to accurately predict the solutions. The PF problem is formulated as a multi-output regression model where two or more output values are predicted based on the inputs. The training and testing data are generated through the OpenDSS-MATLAB COM interface. These methods are completely data-driven where the training relies on reducing the mismatch at each node without the need for the knowledge of the system. The novelty of the proposed methodology is that the models can accurately predict the PF solutions for the unbalanced distribution grids with mutual coupling and are robust to different R/X ratios, topology changes as well as generation and load variability introduced by the integration of distributed energy resources (DERs) and electric vehicles (EVs). To test the efficacy of the DNN models, they are applied to IEEE 4-node and 123-node test cases, and the American Electric Power (AEP) feeder model. The PF results for RBFnet, MLP, and CNN models are discussed in this paper which demonstrate that all three DNN models provide highly accurate results in predicting PF solutions.  
\end{abstract}

\begin{IEEEkeywords}
Power flow analysis, Deep neural networks, Radial basis function networks, Multi-layer perceptron, Convolutional Neural Networks, Unbalanced power distribution grids.
\vspace{-4mm}
\end{IEEEkeywords}

\maketitle

\section{Introduction}\label{Introduction}
\IEEEPARstart{P}{ower} flow (PF) calculation as a fundamental analysis tool for efficient, reliable, and stable operation of the power grid has found widespread use in power systems. PF is extensively used for electrical network analysis and prediction, potential problem identification, determination of the steady-state conditions of the systems, and optimizing the performance of the power grids \cite{pfapplication1, pfapplication2, mirzapour2022, pfapplication3}. The resource mix of distribution systems has been changing due to the increasing penetration of distributed energy resources (DERs) such as wind turbines, solar panels, and energy storage devices \cite{EV_effect, radial_ring}. This transformation necessitates the development of new PF calculation methods to accurately model the uncertainties of these DERs and ensure an efficient and safe operation of the power grid.

Traditional PF calculation methods such as Newton-Raphson and Gauss-Seidel were mostly performed based on radial distribution system structure assuming the unidirectional flow of power from the substation to the consumers  \cite{radial1, radial2, radial3}. The main drawbacks of these models are inaccurate modeling or lack of DERs, limited ability to handle voltage regulation, inability to control power factor, and vulnerability to system reconfigurations. The emergence of inverter-based DERs along with the increasing use of electric vehicles (EVs) have led to a situation where power flow is no longer in a one-way direction, rendering the conventional methods inadequate. {Moreover, the traditional models need exact modeling of the distribution systems from the generation to the load sides and any changes in the system conditions including the topology changes need updated mathematical models for power flow calculations. Additionally, it is not always guaranteed that these models converge to the global optimum and there is a need to develop new optimization methods for every small modification of the grid configurations \cite{PF_GNN_hansen}}. The presence of these intermittent resources in the distribution grids brings new hurdles for the distribution system operators to develop precise models to improve the accuracy of calculations for voltage and current phasors, and power flows, and ensure reliability in medium/low-voltage networks with high R/X ratios. To cope with the associated challenges of nonlinear PF formulations, numerous techniques for simplifying the problem have been developed. Authors in \cite{tradition1} developed a linearized model to generalize the current magnitudes and proposed a linear transformation to mitigate the error of voltage magnitude approximations in solving the AC-OPF problem for an unbalanced distribution network. In \cite{tradition2}, a linearized approximation of active and reactive power of loads in the PQ buses is proposed to simplify the nonlinear PF equations which can be applied to the typical power line impedances and network configurations. The work in \cite{tradition3} used a current injection method to solve the PF problem by creating a linear approximation of the system's equations using a constant coefficient matrix at every iteration. Both works in \cite{tradition4} and \cite{tradition5} proposed a linear power flow calculation model to address the challenges of nonlinearity issues in the equations. While the former modeled the generated reactive power of distributed generation (DG) units as a function of voltage magnitude, which was then combined with a linear approximation of PF equations, the latter proposed linear models for line currents, voltage magnitudes, flowing power, and voltage phasors to reduce the computational costs. These simplifications disregard the non-linear nature of the PF mapping relationships resulting in inaccurate or low-accurate electrical models. The performance of these methods would decline in realistic scenarios with a high rate of power fluctuations.

Recently, with the advancement of technology, widespread deployment of phasor measurement units (PMUs) and smart meters have enabled system operators and utilities to continuously monitor the grid to identify potential problems and improve the system's efficiency and reliability. A wealth of data measured by these devices has embraced the utilization of deep neural networks (DNNs) and other machine learning (ML) methods in various applications in power system data analytics such as load forecasting \cite{energy_LF, short-term_LF}, demand-side modeling \cite{two-sided, iterative}, wind and solar forecasting \cite{solar_battery, solar_radiation}, state estimation \cite{SE_cyber, SE_application}, cyber-power anomaly detection \cite{PIConvAE, PIML}, security assessment \cite{security_mousavi}, and PF calculations \cite{pfmlsurvey}. In particular, these methods are applied to power systems to achieve more accurate and reliable power flow calculation methods for more efficient power grid operations \cite{pfmlsurvey, pfphysics, pfmlarx, DeepOPF, pfDeepRL}. {A detailed discussion of the related works of ML algorithms for PF calculation is provided in Section \ref{Related_work}}. The key advantage of ML algorithms is that they can learn the load-to-solution mappings to successfully capture the nonlinearities in the PF problem while the operating points of the grid change. {The development of the neural network models allows us to achieve more accurate and efficient results compared to traditional methods \cite{PF_GNN_hansen}. This is because the traditional numerical methods assume balanced systems and require the detailed mathematical modeling of the systems which is difficult to obtain in unbalanced power distribution systems in the presence of distributed energy resources. The neural network models can learn the hidden patterns and relationships in the input data to have better approximations for the power flow solutions without the need for exact modeling of the distribution grids and complex system structures. As the NN models work with input-output relationships, they can be scalable and adaptable to larger and more diverse systems without a substantial increase in the computational cost.} However, the main challenge of ML methods especially DNNs lies in ensuring that the generated solutions satisfy power balance equations as well as the operational constraints with minimum error. {Therefore, developing robust and efficient NN models is a need for unbalanced power distribution grids integrated with DERs.}

In this paper, we propose three DNN algorithms; Radial Basis Function Network (RBFnet), Multi-Layer Perceptron (MLP), and Convolutional Neural Network (CNN) to find PF solutions in unbalanced power distribution systems \cite{deepak_thesis}. More specifically, the main contributions of this work can be summarized as follows.
\begin{itemize}
\item{Three DNN models are proposed to determine PF solutions in power distribution systems with unbalanced three-phase configurations and mutual couplings. More specifically, our methods are capable of predicting various variables including branch currents, node voltage magnitudes and angles, current magnitudes and angles, as well as power losses of lines with high precision and very low error rates. The proposed DNN-based methods can be a promising alternative to the traditional iterative-based load flow techniques.}
\item{In addition to the high accuracy of the predicted solutions, the proposed methods are computationally efficient and provide very accurate solutions under different R/X ratios, topology change as well as generation and load variability.}
\item{The efficacy of these approaches is evaluated on the IEEE 4-node system, IEEE 123-node test case, and American Electric Power (AEP) feeder model. We integrate residential and commercial charging stations of EVs and residential rooftop PVs in the IEEE 123-node test case to model the uncertainties of these resources which makes our models more applicable to real-world applications.}
\end{itemize}

The rest of the paper is organized as follows. Section \ref{Related_work} provides a detailed literature review of the related works for predicting PF solutions using ML methods. Section \ref{Power Flow} briefly discusses the load flow used in power distribution systems. Section \ref{proposed methodology} covers the role of big data in power systems and preprocessing of data for our application.  Section VI presents the results and finally, Section VII concludes the paper.    

\section{Related Work} \label{Related_work}
With the fast growth of inverter-based DERs and ongoing progress and complexity of power distribution systems, data-driven methods especially ML have been increasingly utilized to determine PF solutions and find more accurate predictions than traditional methods. Authors in \cite{multi-MG} proposed a probabilistic power flow approach for multi-microgrids to model the microgrids based on the limited information of the distribution systems using a recurrent neural network (RNN) and finding the correlation between input data samples. To adapt the nonlinearity of PF, the nonlinear relationships are converted to linear mapping in \cite{lifting_LR} using the Koopman operator theory and lifting-dimension function. The proposed data-driven method showed better performance in mesh and radial networks. In \cite{GNN_realistic}, an unsupervised graph neural network is presented to capture the topology configuration and physical properties of distribution grids for PF calculations. Authors in \cite{support_matrix} propose a support matrix regression to construct the PF mapping by learning the physical embedding in the observable area and general approximation for unobservable regions. The method also uses the concept of margins to be robust against outliers. To solve the distribution grid optimization, the work in \cite{Nonlinearity_Adaptive} presents a dimension lifting data method for mapping power PF results (power injection to voltage magnitude mapping) while considering the constraints of DERs.
The effects of outliers such as leverage points and vertical outliers are considered in \cite{Data-driven_RBM} to develop PF analysis based on a data-driven robust process model (RBM) which can handle a significant portion of outliers in voltage phasor and power injection measurements. Authors in \cite{robust_mapping} proposed a support vector regression (SVR) approach to recover the main relationships in the PF equations. The proposed method can estimate the mapping rules between voltage phasors and power injections given a set of real power injections, voltage phasors, and reactive power injections.  

{The ability of graph convolution neural network (GCNN) for optimal PF calculation is showcased in \cite{OPF_gao_GNN}. The GCNN model can embed the topology of the grid to extract essential features from local neighborhoods and formulate the loss function to achieve high accuracy in optimal power flow (OPF) calculation under topology changes. The authors in \cite{PI_opf_nellikkath} integrated the OPF equations and constraints in the form of the AC optimal OPF Karush–Kuhn–Tucker (KKT) conditions in the training process of the NN models to determine the optimal parameters for OPF calculations. The proposed framework can the worst-case violations of the NN models by following the OPF equations. A decentralized ML model, Alternating Direction Method of Multipliers (ADMM), was proposed in \cite{decentralized_ADMM} to learn the primal and dual consensus variables to solve AC OPF. The model applied data-filtering techniques to distinguish high-quality data for training and improving the model's accuracy. The researchers in \cite{PF_UC_alyssa} developed an optimally compact NN model by transforming the nonlinearities of PF equations into piecewise
linear approximations using mixed integer linear programs (MILPs). The proposed strategy allows for maintaining a high number of binary variables that are used for unit commitment. The model presented in \cite{typed_GNN} integrated the information of grid branches and different bus types into a graph representation to solve the AC PF problem. The presented method can identify the grid topology changes and is applicable for line outage scenarios. A Bayesian regularized Deep Neural Network (BDNN) is proposed in \cite{bayesian_DCMG} to reduce the loss in load expectation and assess the voltage quality and various reliability indexes by solving the power flow problem in a DC microgrid}
 
In this paper, we aim to present trained neural networks that predict branch currents, node voltages, power losses, and voltage and current angles, with very low error margins in an unbalanced power distribution grid with mutual couplings. Our trained deep neural networks can serve as a substitute for the traditional iterative-based load flow techniques. In addition, it can replace the conventional test cases provided by utilities for research purposes. 
 
\section{Distribution Grid Power Flow} \label{Power Flow}

Power flow for distribution systems is different than transmission system power flow due to factors such as the high R/X ratio of feeders, the radial structure of the system, and unbalanced generation and loads.
Load flow techniques for transmission systems like Gauss-Siedel and Newton-Raphson are not efficient and do not converge to the optimal solutions for radial distribution systems \cite{thukaram1999robust}.

\begin{table}[]
\caption{\small FBS summary \cite{FBS_solanki}.}
\label{FBS Table}
\centering
\scalebox{0.75}{
\begin{tabular}{|l|ll|l|}
\hline
\multicolumn{1}{|c|}{\textbf{Component}} &  & \multicolumn{1}{c|}{\textbf{Backward Sweep}} & \multicolumn{1}{c|}{\textbf{Forward Sweep}} \\ \hline
\multirow{2}{*}{\begin{tabular}[c]{@{}l@{}}Overhead Lines/\\ Underground cables\end{tabular}} & \multicolumn{2}{l|}{$V_i^{new} = V_{spec} -ZI_i$ for first calculation} & $V_i = V_{i+1}+ZI_{i+1}$ \\
 & \multicolumn{2}{l|}{$V_{i+1} = V_i^{new}-ZI_{i+1}$} & $I_i=I_{i+1}$ \\ \hline
 \multirow{2}{*}{\begin{tabular}[c]{@{}l@{}}Three Phase\\ Transformer\end{tabular}} & \multicolumn{2}{l|}{$V_{i+1} = A_tV_i^{new}-B_tI_{i+1}$} & {$V_i = a_tV_{i+1}+b_tI_{i+1}$} \\
 & \multicolumn{2}{l|}{ } & $I_i=d_tI_{i+1}$ \\ \hline
\multirow{2}{*}{Switch} & \multicolumn{2}{l|}{$V_i^{new}=K*V_{spec}$} & $V_i=K*V_{i+1}$ \\
 & \multicolumn{2}{l|}{$V_{i+1}=K*V_i^{new}$} & $I_i=K*I_{i+1}$ \\ \hline
\multirow{2}{*}{Capacitor} & \multicolumn{3}{l|}{$Z_i^{ph-n}=\frac{\lvert V_i^{ph-n}\rvert^2}{S_i^{ph-n*}}$,$IC_i^{ph-n}=\frac{V_i^{ph-n}}{Z_i^{ph-n}}$ Wye Connected} \\
 & \multicolumn{3}{l|}{$Z_i^{ph-ph}=\frac{\lvert V_i^{ph-ph}\rvert^2}{S_i^{ph-ph*}}$,$IC_i^{ph-ph}=\frac{V_i^{ph-ph}}{Z_i^{ph-ph}}$ Delta Connected} \\ \hline
\multirow{6}{*}{Load} & \multicolumn{1}{c|}{Load Type} & \multicolumn{1}{c|}{Wye} & \multicolumn{1}{c|}{Delta} \\ \cline{2-4} 
 & \multicolumn{1}{l|}{Const PQ} & $IL_i^{ph} = \left(\frac{S_i^{ph-n}}{V_i^{ph-n}}\right)^*$ & $IL_i^{ph} = \left(\frac{S_i^{ph-ph}}{V_i^{ph-ph}}\right)^*$ \\ \cline{2-4} 
 & \multicolumn{1}{l|}{\multirow{2}{*}{Const Z}} & $Z_i^{ph-n}=\frac{\lvert V_i^{ph-n}\rvert^2}{S_i^{ph-n*}}$ & $Z_i^{ph-ph}=\frac{\lvert V_i^{ph-ph}\rvert^2}{S_i^{ph-ph*}}$ \\
 & \multicolumn{1}{l|}{} & $IL_i^{ph-n}=\frac{V_i^{ph-n}}{Z_i^{ph-n}}$ & $IL_i^{ph-ph}=\frac{V_i^{ph-ph}}{Z_i^{ph-ph}}$ \\ \cline{2-4} 
 & \multicolumn{1}{l|}{\multirow{2}{*}{Const I}} & $IL_i^{ph-n}=\lvert IL_i^{ph-n}\rvert^*$ & $Z_i^{ph-ph}=\frac{\lvert V_i^{ph-ph}\rvert^2}{S_i^{ph-ph*}}$ \\
 & \multicolumn{1}{l|}{} & $\angle \delta_i^{ph-n} - \theta_i^{ph-n} $ & $IC_i^{ph-ph}=\frac{V_i^{ph-ph}}{Z_i^{ph-ph}}$ \\ \hline
\multirow{2}{*}{Distributed Load} & \multicolumn{3}{l|}{\begin{tabular}[c]{@{}l@{}}1. Two-thirds load is lumped at one-fourth length of the line from \\ sending end\end{tabular}} \\
 & \multicolumn{3}{l|}{2. One-third load is lumped at the receiving node.} \\ \hline
 
\end{tabular}}
\end{table}

The forward-backward sweep (FBS) method is one of the most commonly used distribution system power flow (DSLF) methods \cite{kersting2017distribution}. The fast decoupled power flow method is also used for radial distribution power flow \cite{zimmerman1995fast}. \cite{teng2003direct} uses bus injection to branch current (BIBC) matrix and branch current to bus voltage (BCBV) matrix to solve bus voltages iteratively.

The detailed model for the FBS method is presented in detail in Table \ref{FBS Table}. In this work, we have used OpenDSS to perform load flow and generate training and testing data (explained in subsequent sections). OpenDSS uses the solution of the nonlinear admittance equation to analyze the load flow as mentioned in \cite{dugan2016reference}: $I(V)=Y_{system}V$, where $I$ is the injection current from power conversion elements such as load, PV, storage, and generators in the circuit and $Y_{system}$ is the admittance matrix of the system. The solution of this equation is found to work well for the distribution network if the initial voltage is close to the final solution.

The unbalanced three-phase PF equations are represented by considering the mutual inductance and inter-phase capacitance among different phases\cite{linear_PF}. The equations can be given as follows.

\begin{align}
  {P^r_i}
  = {V^r_i}\sum_{k\in i, s=a,b,c}{V^s_k} \times
  \mspace{150mu}
  \notag\\
     ({G^{ik}_{rs}}\cos(\theta^r_i-\theta^s_k)+{B^{ik}_{rs}}\sin(\theta^r_i-\theta^s_k))
\vspace{-2mm}
\end{align}
\begin{align}
\vspace{2mm}
  {Q^r_i}
  = {V^r_i}\sum_{k\in i, s=a,b,c}{V^s_k} \times
  \mspace{150mu}
  \notag\\
     ({G^{ik}_{rs}}\sin(\theta^r_i-\theta^s_k)-{B^{ik}_{rs}}\cos(\theta^r_i-\theta^s_k))
\end{align}
where $i$ and $k$ are node indexes, $r$ and $s$ denote phase indexes, $P^r_i$ and $Q^r_i$ are active and reactive power injections of phase $r$ and node $i$, $\theta^r_i$ and $\theta^s_k$ represent the phase angle of  phase $r$ in node $i$ and phase $s$ in node $k$,  and $V^r_i$ denotes the voltage magnitude and phase angle of phase $r$ and node $i$. $G^{ik}_{rs}$ and $B^{ik}_{rs}$ are $3\times3$ matrices denoting conductance and susceptance between phases $r$ and $s$, and node $i$ and $k$ which comprise the mutual admittance matrix $Y^{ik}$; ${Y^{ik}} = {G^{ik}} + j {B^{ik}}$.

Load modeling in unbalanced distribution grids is different than transmission systems. Due to load natures in distribution grids, the nodal net injected current $I$ is defined as a function of nodal voltages $V$. Generally, loads are divided into three categories as mentioned in Table \ref{FBS Table}. For the ZIP load model, the nodal current for each phase $ph$ in node $i$ is defined by the following function
  \begin{equation}
I^{ph}_{i}(V_i)=I^{ph}_{I_i}(V_i) + I^{ph}_{PQ_i}(V_i) + I^{ph}_{Z_i}(V_i) 
\label{MinMax}
\end{equation}
where $I^{ph}_{I_i}(V_i)$, $I^{ph}_{PQ_i}(V_i)$, and $I^{ph}_{Z_i}(V_i)$ are currents from constant-current loads, constant-power loads, and constant-impedance loads, respectively. These currents are defined for the Wye and Delta connection in Table \ref{FBS Table}.
 
\section{Proposed Methodology} \label{proposed methodology} 
In this section, the principles of the three NN models, i.e., MLP, RBFnet, and CNN, are thoroughly explained and then, the performance metrics are introduced. The schematic diagram of our proposed technique is shown in Fig. \ref{Outline}. As shown in this figure, the training data is generated by load flow analysis using the OpenDSS-MATLAB COM interface for our test cases. The advantage of OpenDSS is that any type of distribution grid with various topologies can be modeled through writing transcripts. Using the COM interface, OpenDSS can interchange data with other tools like MATLAB and provides better flexibility to users to modify the source code and perform the objective.  

The generated datasets from the OpenDSS-MATLAB COM interface are the load flow quantities such as node voltage magnitudes and angles, branch currents, and node current magnitudes and angles. These values along with the source voltage, line parameters, and loadshapes are fed to NN models for training. The training data is normalized through min-max normalization prior to model training since all the bus voltages and currents differ in range.   

\subsection{Neural Network Models} \label{NN Models}

\begin{figure}[]
  \begin{center}
\includegraphics[width=0.5\textwidth]{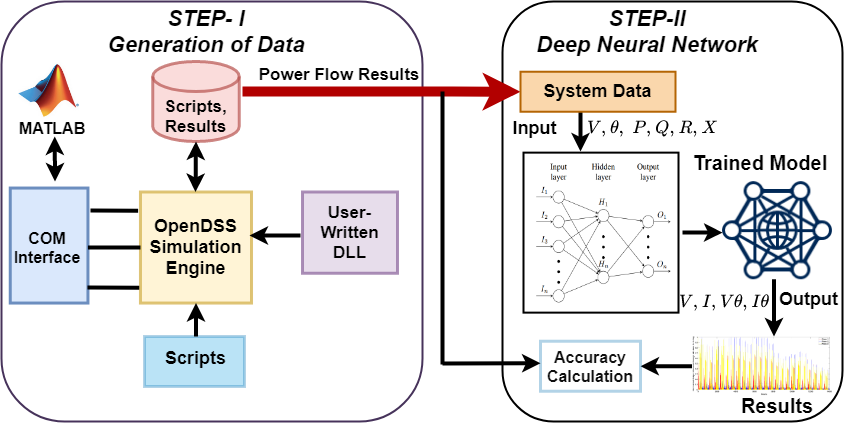}
\caption {\small Outline of Proposed Methodology} 
\label{Outline}
\vspace{-4mm}
\end{center}
 \end{figure}

As shown in Fig. \ref{Outline}, the input data to NN models at $i$-th timestep are the generated output data of OpenDSS expressed as
\begin{equation}\label{input}
 X_i = [\mid V_i \mid, \theta_{V_i}, Z, P, Q]
\end{equation}\
where $\mid V_i \mid$ represents voltage magnitudes, $\theta_{V_i}$ denotes voltage angles, $Z$ is the impedance of the system, and $P$ and $Q$ represent the active and reactive powers of the loads in the system, respectively.
After feeding these inputs to NN models, they are trained to predict the PF solutions which are the voltage and current magnitudes as well as voltage and current angles of the system's nodes and line power losses as represented below.
\begin{equation}\label{output}
 X_o = [\mid V_i \mid, \mid I_i \mid, \theta_{V_i}, \theta_{I_i}, P_{l_{ij}}]
\end{equation}\
{The voltage magnitudes and angles in (\ref{input}) represent the outputs of the openDSS simulation which act as the initial voltage magnitudes and angles that along with the system information (load characteristics and line impedances) can be applied to the neural network models for calculation of the power flow solutions. This would enable the model to learn the relationships between the given system conditions for better predictions. These initial values typically are obtained using real-time measurements, historical data, or power system simulations. On the other side, the voltage magnitudes and phase angles in (\ref{output}) are the predicted values after training the neural network models for a certain number of epochs.}

We have employed three different NN models for predicting PF solutions: (1) multi-layer perceptron (MLP); (2) convolutional neural network (CNN); and (3) radial-basis function (RBF) network. In the following, we discuss the general details of these NN models, and the specific implementation details for our proposed methods are provided in Section \ref{Results}.  

\subsubsection{\textbf{Multi-layer Perceptron}}

A perceptron or a neuron is a linear classifier i.e., it classifies inputs by separating two categories with a straight line. Inputs are typically represented by a feature vector $x$, which are multiplied by a set of learnable weights $w$ and added to a bias $b$: $y = w * x + b$. A perceptron produces a single output based on several real-valued inputs by forming a linear combination using its input weights (and sometimes passing the output through a nonlinear activation function). It is given as:

\begin{equation}
    y=\phi(\sum_{i=1}^{n}w_{i}x_{i}+b)=\phi(\textbf{w}^{T}\textbf{x}+b),\label{eq:4}
\end{equation}where \textbf{w} denotes the vector of weights, \textbf{x} is the vector of inputs, b is the bias, and $\phi$ is the non-linear activation function.

MLP is a finite-directed acyclic graph artificial neural network composed of more than one neuron. Like other NNs, it receives the inputs, passes them through an arbitrary number of hidden layers, and the output layer makes a decision or prediction about the inputs. 
MLPs are often applied to supervised learning problems: they are trained on a set of input-output pairs and learn to model the correlation (or dependencies) between those inputs and outputs. Training involves adjusting the weights and biases of the model in order to minimize the sum of the errors. Backpropagation is used to make those weight and bias adjustments relative to the error, and the error itself can be measured in a variety of ways, including by root mean squared error (RMSE).

MLP is a feed-forward network that consists of two motions: a forward pass and a backward pass. In the forward pass, the input moves from the input layer through the hidden layer to the output layer, where the output layer makes a decision. An error is calculated by comparing the decision at the output layer against the ground truth (GT) labels.
In the backward pass, partial derivatives of the error function with respect to weights and biases are calculated to update the learning parameters through backpropagation. This process can be implemented using any kind of gradient-based optimization such as stochastic gradient descent (SGD) or Adam optimizer. These forward and backward passes are repeated till the error reaches a predefined value or until a certain number of iterations or epochs.  
    
\subsubsection{\textbf{Convolutional Neural Network}}
Convolutional Neural Networks \cite{CNN_introduction} are very similar to MLPs: they are made up of perceptrons (neurons) with learnable weights and biases. The noticeable difference between CNN and MLP is that CNNs are primarily used for inputs of image type which allows us to capture specific patterns of input data.  

CNN consists of three basic components:
convolutional, pooling, and fully connected layers. The purpose of embedding the convolutional layer is to learn a filter bank given input feature maps. The input of a convolution layer is in the form of a $n_1 \times n_2 \times d$, where $d$ denotes the number of two-dimensional input feature maps with the height $n_1$ and width  $n_2$. Let $x_{i}(j,k)$ denote the component at row $j$ and column $k$ in the $i$th feature map, and we use $x_i^{l}$ to denote the complete $i$th feature map at layer $l$. In the learning process of CNNs, $f$ sets of filters of size $k_1 \times k_2$ are learned to output $x_{(l+1)}$ for the next layer. This feature is a three-dimensional array with $f$ number of two-dimensional feature maps of size $(n_1-k_1+1) \times (n_2-k_2+1)$. More formally, the convolution layer computes the following:

\begin{equation}
    x_j^{(l+1)}=s(\sum_i K_{ij}*x_i^{(l)}+b_j),
\end{equation}where $K_{ij}$ denotes the filter that connects the input feature map $x_i^{l}$
to the output feature map $x_j$ , $b_j$ is the bias for the $j$th output feature map, $s(.)$ is elementwise non-linear function
and $\ast$ denotes the discrete two-dimensional convolution. We denote $j$th convolutional layer with $f$ filters of size $k_1 \times k_2$ by $C_{(f,k_1 \times k_2)}^{(j)}$.

It is very common to insert a pooling layer in between the convolutional layers of a CNN. The pooling layer is responsible for reducing the spatial size of the output of the convolutional layer to reduce the amount of parameters and computation in the network, and also control over-fitting. Furthermore, the pooling layer is useful for extracting dominant features that are position and rotation invariant, thus maintaining the process of effectively training the model. 
The pooling layer operates independently on every feature map (output of convolutional layer) and resizes it spatially, using the max or average operation.  For example, with
a pooling filter size of $m_1 \times m_2$ an input with $f$ number of two-dimensional feature maps of size $n_1 \times n_2$ will result in an output of $f$ number of feature maps of size $n_1/m_1 \times n_2/m_2$. For average (mean) pooling, the output will be the average of all the values from the portion of the image covered by the filter, and for max pooling the output will be the maximum value from the portion of the image covered by the filter. We denote the $j$th pooling layer with pooling filter size of $m_1 \times m_2$ as $P^{(j)}_{m_1 \times m_2}$. 

The convolutional layer and the pooling layer, together are considered to be the $i$-th layer of a CNN. Depending on the application requirements and complexities in the images, the number of such layers can be adjusted to acquire the required low-level details. The output of the last pooling layer is flattened and fed to the fully connected layer for further processing. A fully connected layer is usually added to a CNN to learn non-linear combinations of the high-level features generated at the output of the convolutional or pooling layer. 
Neurons in a fully connected layer have full connections to all activations in the previous layer. Their activations can hence be computed with a matrix multiplication followed by a bias offset as given in (\ref{eq:4}). We denote the $j$th fully connected with $h$ neurons or hidden units as $F^{(j)}_h$.  

\subsubsection{\textbf{Radial Basis Function Network}}

Radial basis function network (RBFN) known as kernel function is used for regression or function approximation. RBFNs can be used to model an underlying trend or a function using many Gaussian (bell) curves.
An RBFN resembles a 3-layer MLP network consisting of an input layer, a hidden layer made of radial basis function (RBF) neurons, and an output layer with one node per category or class of data. 
The entire input vector is passed through each RBF neuron. A “prototype” vector, which is one of the vectors from the training set is stored in each RBF neuron. The input vector is compared against the prototype in each RBF neuron to provide an output value between 0 and 1 showing a measure of similarity. The more similarity of the input and the prototype, the closer the output of that RBF neuron to 1. As the distance between the input and prototype grows, the response falls off exponentially towards 0. There are different functions that could be used to measure the similarity. However, in an RBFN, the most commonly used similarity or activation function is RBF, which is based on Gaussian distribution. The Gaussian distribution  is given by:

\begin{equation}
    f(x)=\frac{1}{\sigma\sqrt{2\pi}}e^{\frac{-(x-\mu)^2}{2\sigma^2}},
\end{equation}where x is the input, $\mu$ is the mean and $\sigma$ is the standard deviation. However, for RBF neuron activation, we use the RBF function given as:

\begin{equation}\label{Gaussian_RBF}
    f(x)=e^{\frac{-(x-\mu)^2}{2\sigma^2}}.
\end{equation}

The output layer consists of a set of nodes, {each of which represents a measurement in the output vector $X_o$ in (\ref{output}). That is, each of the output nodes is designed to predict a specific measurement (voltage magnitude, voltage phase angle, current magnitude, current phase angle, or line power loss)}. Each output node computes a sort of score for the associated category. A weighted sum of the activation values from every RBF neuron is taken to compute the category score. By weighted sum, we mean that each RBF neuron is associated with a weight value by the output node. This weight value is multiplied by the neuron's activation before adding it to the total response. Every output node has its own set of weights and the output node will typically give a positive weight to the RBF neurons that belong to the true category and a negative weight to the others.

To summarize, given an input $x$, an RBF network produces a weighted sum output:

\begin{equation} 
F(x)=\displaystyle\sum_{j=1}^k w_j\varphi_j(x, c_j) + b
\end{equation}
where $w_j$ are the weights, b is the bias, k is the number of bases/clusters/centers, $c_j$ represents the $j$th cluster, and $\varphi_j(\cdot)$ is the Gaussian RBF described as in (\ref{Gaussian_RBF}).

The centers of the RBF neurons can be obtained in different ways as long as the entire data space is well represented. These ways are summarized as:

(1). They can be chosen randomly from the training dataset.

(2). The centers could be selected using unsupervised learning techniques such as K-means clustering \cite{duda2012pattern}.

(3). The centers could also be obtained through supervised learning.

{In this paper, we implemented the K-means algorithm as follows to obtain the centers of the RBF neurons. 1) The initial number of clusters $K$ is randomly selected and the clusters are randomly initialized. 2) The distances between each data point and each cluster are computed and each data point is assigned to the closest cluster. 3) The mean of all points in each cluster is computed to update the cluster center. 4) The standard deviation of the clusters is calculated. 5) The clusters with zero or one point are identified and handled by computing a mean standard deviation based on the clusters with sufficient points. 6) If the distance between the current and previous cluster centers is below a threshold (in our case $10^{-6}$), the algorithm is converged. 7) Otherwise, go to step 2. 8) The final cluster centers and the corresponding standard deviations are returned to be used for the RBF network.\\
It is worth mentioning that in step 5 since the clusters with insufficient data points are unreliable and may lead to inaccurate cluster centers or standard deviations, they are tracked to provide more robust estimations of the cluster's characteristics. After the convergence, the returned clusters are used as the centers of the RBF neurons.} 

\section{Experimental Results}\label{Results}
In this section, the performance of the three proposed DNNs in finding PF solutions is extensively evaluated by implementing them for different case studies as explained in \cite{deepak_thesis}. Three test cases including the IEEE 4-bus system, AEP test feeder, and IEEE 123-bus feeder are simulated to show the efficacy of the proposed DNN methods for predicting PF solutions of three-phase unbalanced power distribution grids. Additionally, to show the applicability of the methods in real-world applications, they are evaluated under topology changes and with the integration of EVs and residential rooftop PVs in the IEEE 123-node test case. To calculate the prediction accuracy and evaluate the efficacy of the trained models in finding the PF solutions, mean absolute error (MAE) and mean square error (MSE) are employed.

\subsection{Case 1: IEEE 4 Node Test Case}
\begin{figure}[h]
    \centering
    \includegraphics[clip, trim=-1.03cm 0cm -1.1cm 0cm, width=0.5\textwidth]{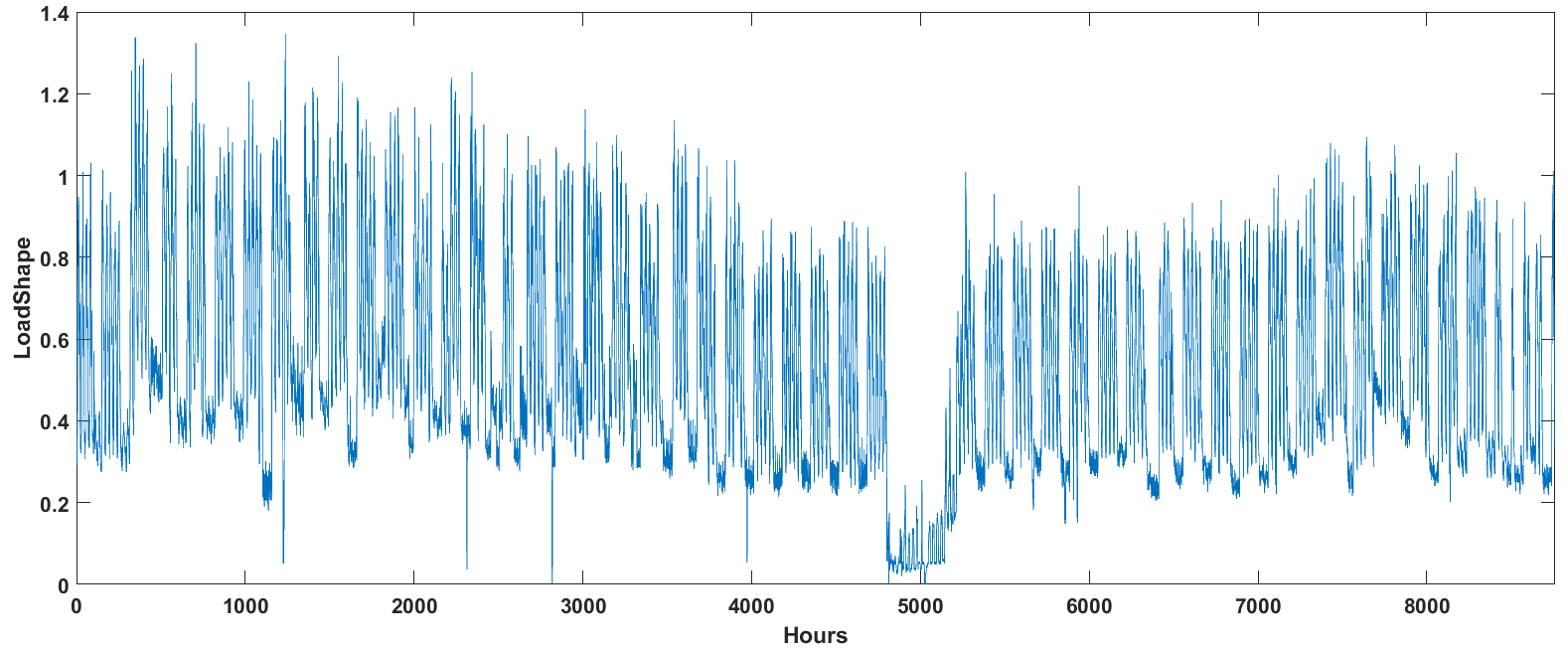}
    \begin{center}
    \caption {\small Loadshape for load at bus 4.} 
    \label{fig:Loadshape}
    \end{center}
    \vspace{-2mm}
     \end{figure}

Fig. \ref{fig:Loadshape} refers to the loadshape of the load at node 4 of the IEEE 4-node distribution network. We have generated the data for 3 years using the OpenDSS platform. Therefore, the total data points are equal to 26280. Among them, 21000 points are randomly selected for training of the neural networks and 5280 points for testing and analysis. Considering the single load in the IEEE 4-node distribution network, the input vector dimension for the NNs is equal to 43 {which includes the entries of node voltage magnitudes, voltage phase angles, active and reactive load, and line impedances for the three-phase unbalanced network,} and the number of classes or outputs is equal to 55 {which includes node voltage magnitudes and phase angles, node current magnitudes and angles, branch currents, and line power losses of the three-phase system}.    

The implementation and the network parameters for the three proposed models are given in the following:
\begin{enumerate}
    \item \textbf{Convolutional neural network}:  For the CNN model training and testing, the size of the input vector for all the data points is reduced to 42 by discarding a single value from the X Matrix. The reduced input vector of size 42 is reshaped into $7 \times 6$ to make it possible to be used as input for CNN. We just use a single layer of input vector $7 \times 6$ instead of the traditional three-layer input for CNN. The training is carried out by optimizing the MSE objective using mini-batch gradient descent with momentum. 
      
    The architecture that we used for CNN is as follows: two convolutional layers of 32 filters of size $2 \times 2$ and 64 filters of size $2 \times 2$. We have used a max pooling layer of size $2\times 2$ only for the first convolutional layer. The convolutional and pooling layers are followed by four fully connected layers of size $512, 256, 128$, and $64$ in that order, and finally the output layer of size $54$. We use the rectified linear unit (ReLU) activation function for all layers except the last layer. We have not used any activation for the last layer. The optimization is implemented by minimizing the MSE of the actual and predicted values using the Adam optimizer with a momentum of 0.5. The batch size is set to 256 with the learning rate equal to 0.0001. We have trained the CNN for 1000 epochs. 
    \item \textbf{Multi-layer perceptron}: For MLP, the size of the input vector is equal to 43 with no value being discarded. For the MLP model, we have used four hidden fully connected layers of size $512, 256, 128$, and $64$ in that order. The output layer is of size $54$. The rest of the training details are consistent with the CNN model except for the number of training epochs. The MLP model has been trained for 100 epochs.
    \item \textbf{Radial-basis function network}: Similar to MLP, even for the RBFN model, the size of the input vector is equal to 43. We have used only one hidden RBF layer. The number of RBF neurons in the hidden layer is equal to 50. We have used K-means clustering to calculate the centers of the RBF neurons. Similar to CNN and MLP, the MSE loss minimization is implemented through Adam optimizer with a momentum of 0.5. The batch size was set to 1 with the learning rate equal to 0.0001. We have trained the RBFN for 150 epochs.    
\end{enumerate}

\begin{figure}[h]
\vspace{-8mm}
    \centering
       \includegraphics[width=0.42\textwidth]{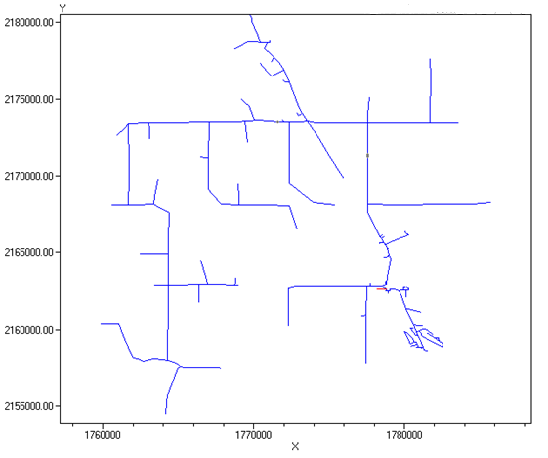}
    \hfill
  \caption{\small AEP feeder}
  \label{fig:AEP Feeder}
  \vspace{-3mm}
\end{figure}

\subsection{Case 2: AEP Distribution System}

The AEP feeder, shown in Figure \ref{fig:AEP Feeder}, represents a large distribution system provided by American Electric Power utility for study purposes. For the AEP distribution network, we have randomly selected 7200 points for training of the NN models and 1560 points for testing and analysis. Considering the AEP distribution network, the input vector dimension for the NN is equal to 2809 {which includes the entries of node voltage magnitudes, voltage phase angles, active and reactive loads, and line impedances for the three-phase unbalanced network,} and the number of classes or the outputs is equal to 5880 {which includes node voltage magnitudes and phase angles, node current magnitudes and phase angles, branch currents, and line power losses of the three-phase system}.    

The implementation and the network parameters for each NN model are given below:
\begin{enumerate}
    \item \textbf{Convolutional neural network}:  For the CNN model in AEP, the input vector size is 2809, which is reshaped into $53 \times 53$ to make it possible to be used as input for CNN. The training is carried out by optimizing the MSE objective using mini-batch gradient descent with momentum. 
    
    The architecture that we use for CNN is as follows: two convolutional layers of 64 filters with size $2 \times 2$, and 128 filters with size $2 \times 2$, followed by two more convolutional layers with 256 filters with size $2 \times 2$. We have used a max pooling layer of size $2\times 2$ for all the convolutional layers. The convolutional and pooling layers are followed by four fully connected layers of size $4096, 2048, 2048$, and $1024$ and finally the output layer of size $5880$. The remaining hyperparameters of the AEP CNN model are consistent with the 4-node CNN model. For example, in the CNN model for AEP, we have used the ReLU activation function for all layers except the output layer and have not used any activation for the output layer, which is very similar to the 4-node model. Similarly, we have used MSE loss as our objective function. Additionally, we have used the Adam optimizer with a momentum of 0.5. The batch size is set to 48 with the learning rate equal to 0.0008. We have trained the CNN for 1200 epochs. The CNN sequence for the AEP distribution is shown in Fig. \ref{fig:cnn_aep}.
    
\begin{figure}[!ht]
    \includegraphics[width=8.5cm]{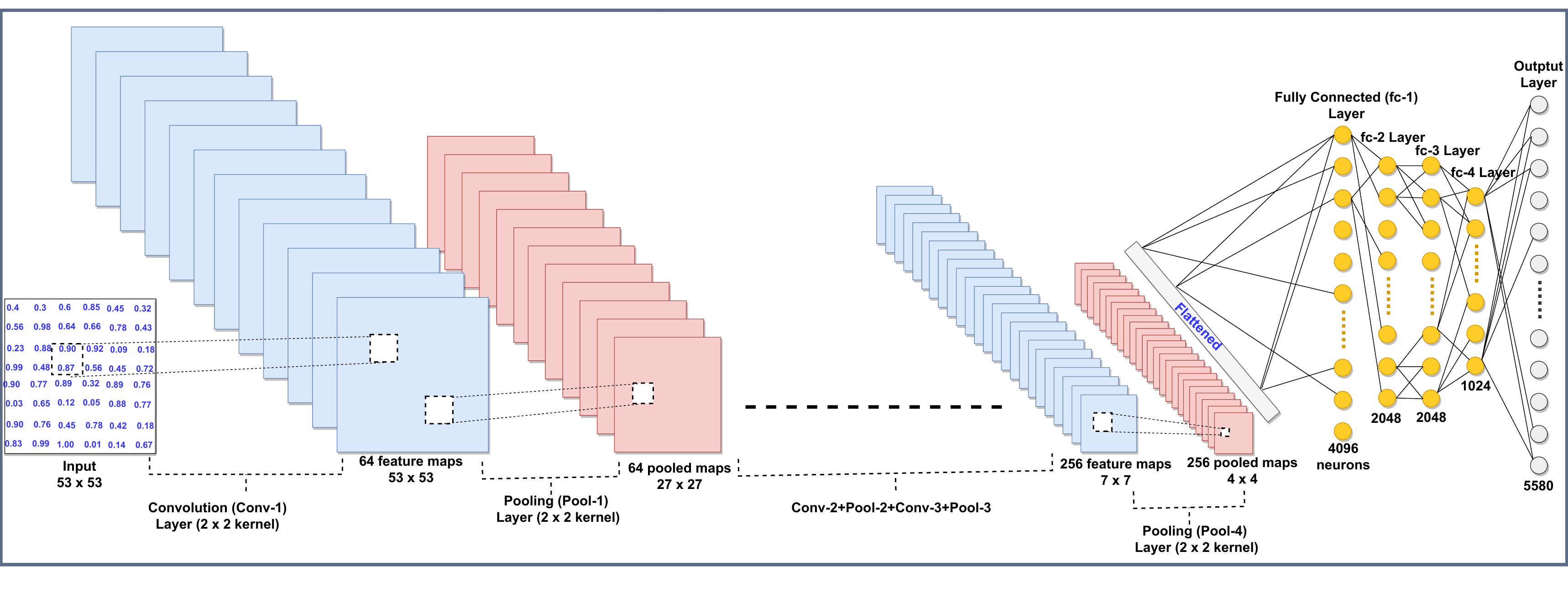}
    \begin{center}
    \caption {\small CNN sequence for AEP distribution. For clarity reasons, all the connections of the fully connected layers have not been shown. The black dashed line indicates that there are more convolutional layers.  } 
    \label{fig:cnn_aep}
    \end{center}
    \vspace{-3mm}
\end{figure}

    \item \textbf{Multi-layer perceptron}: The size of the input vector in the MLP architecture for AEP is equal to 2809 with no value being discarded. The MLP architecture for AEP is constructed by using five hidden fully connected layers of size $4096, 2048, 2048, 1024$, and $1024$. The output layer is of size $5880$. We have used the Adam optimizer with a momentum of 0.5. The batch size was set to 96 with the learning rate equal to 0.0008. We have trained the MLP for 1200 epochs.
    \item \textbf{Radial-basis function network}: Similar to CNN and MLP, the size of the input vector for the RBF network is equal to 2809. We have used only one hidden RBF layer. The number of RBF neurons in the hidden layer is 3500. We have used K-means clustering to calculate the centers of the RBF neurons. We have minimized the mean square error loss for the output layer. The batch size was set to 1 with the learning rate equal to 0.0001. We have trained the RBFN for 150 epochs.    
\end{enumerate}
\begin{table*}[ht]
     \centering
     \caption{\small Results}
     \scalebox{1.3}{
     \begin{tabular}{|c |c|c|c|c|c|c| c|} 
     \hline
     Test Cases & Scenarios & \multicolumn{2}{c|}{RBFnet}& \multicolumn{2}{c|}{MLP}& \multicolumn{2}{c|}{CNN}\\
      \hline
     &   & MSE & MAE & MSE & MAE & MSE & MAE\\
         \hline
     \multirow{3}{*}{IEEE 4 Node} &  Constant PQ Load & 0.12 $\%$ & 0.33 $\%$ & 0.08 $\%$  & 0.36$\%$ & 0.07 $\%$   & 0.35 $\%$ \\ \cline{2-8}

                          &   Constant Impedance Load & 0.11 $\%$ & 0.33 $\%$ & 0.08 $\%$  & 0.36 $\%$ & 0.06 $\%$ & 0.35 $\%$\\\cline{2-8}
      
                          &  ZIP Load & 0.12 $\%$ & 0.32 $\%$ & 0.19 $\%$ & 0.95 $\%$ & 0.06 $\%$ & 0.36 $\%$  \\
        \hline
        
        AEP Feeder & Constant PQ Load &  0.11 $\%$ &  0.34 $\%$ & 0.042 $\%$ & 0.43 $\%$ & 0.04 $\%$  & 0.42 $\%$\\
        \hline
        IEEE 123 Node & Constant PQ Load &  0.10 $\%$ &  0.31 $\%$ & 0.60 $\%$ & 0.10 $\%$ & 0.60 $\%$  & 0.10 $\%$\\
        \hline
        
     \end{tabular}}
     
     \label{tab:my_label}
 \end{table*}

\begin{figure}[t!]
    \subfloat[Normalized output voltage]{%
        \includegraphics[width=.54\linewidth]{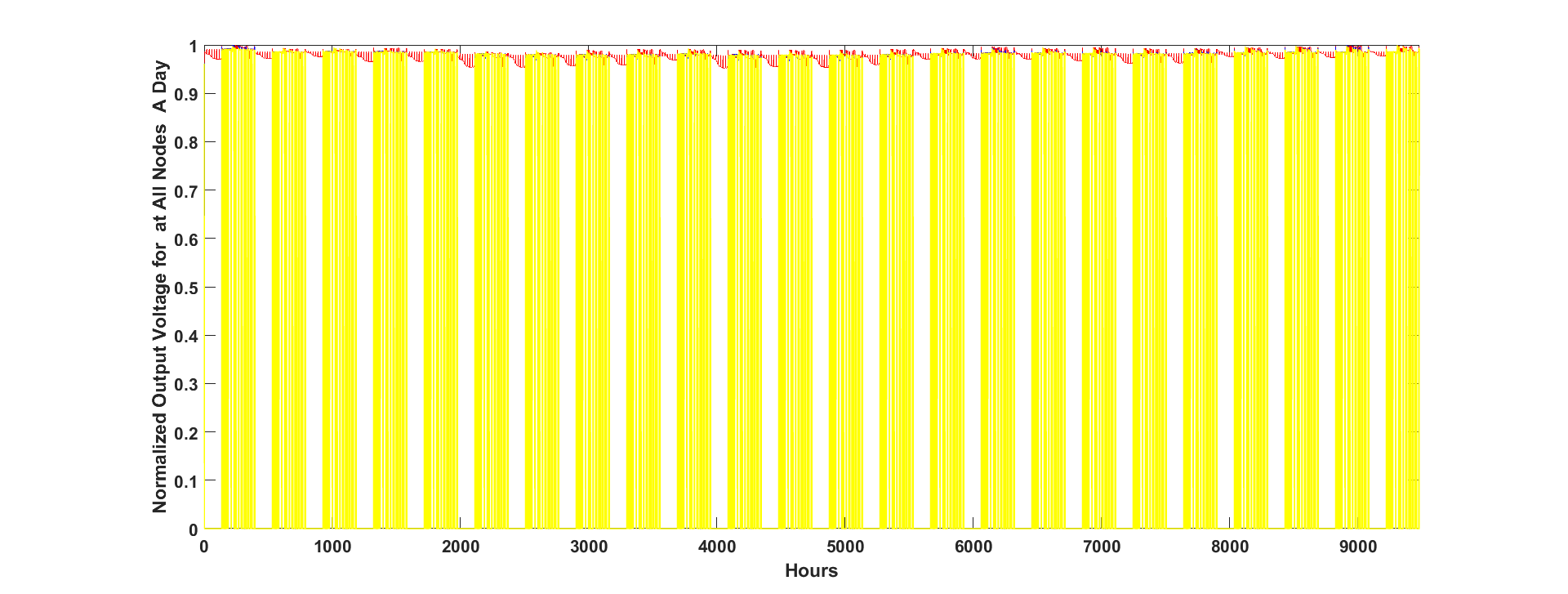}%
        \label{volt_mag}%
    }
    \subfloat[Normalized output current]{%
        \includegraphics[width=.54\linewidth]{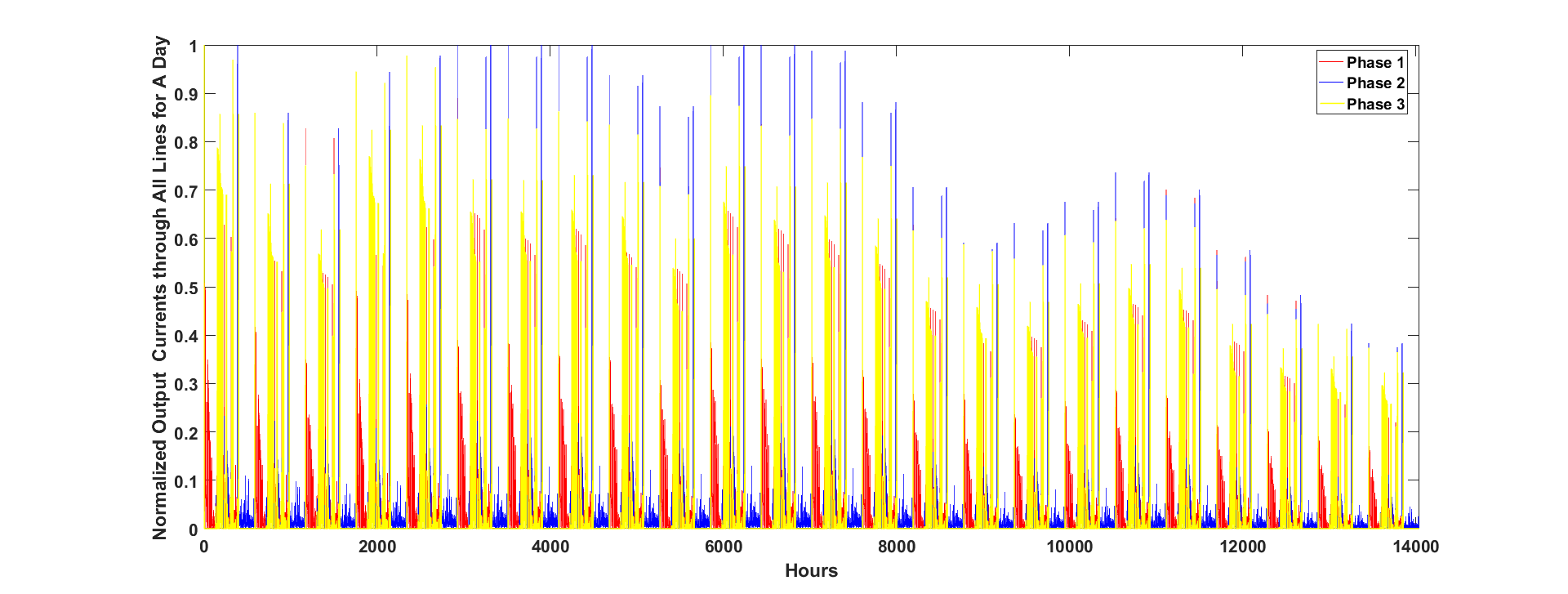}%
        \label{curr_mag}%
    }\\
    \subfloat[Normalized output voltage angle]{%
        \includegraphics[width=.54\linewidth]{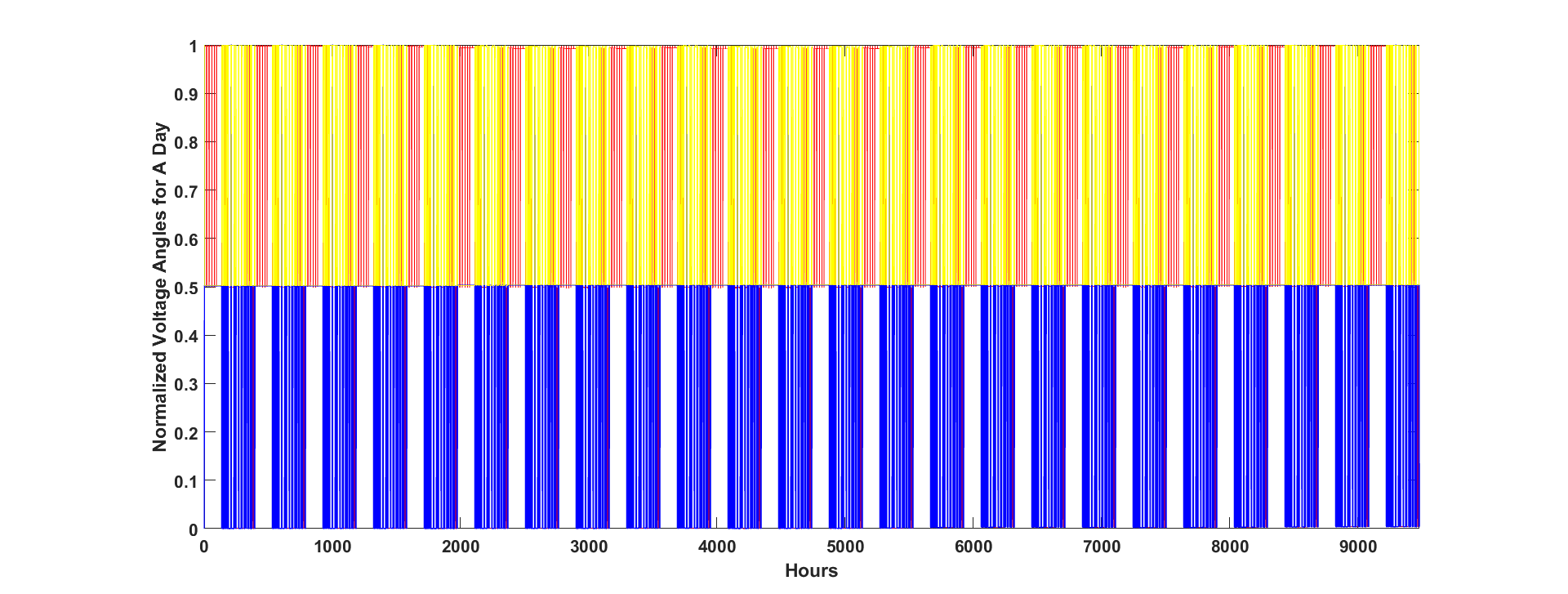}%
        \label{volt_ang}%
    }
    \subfloat[Normalized output current angle]{%
        \includegraphics[width=.54\linewidth]{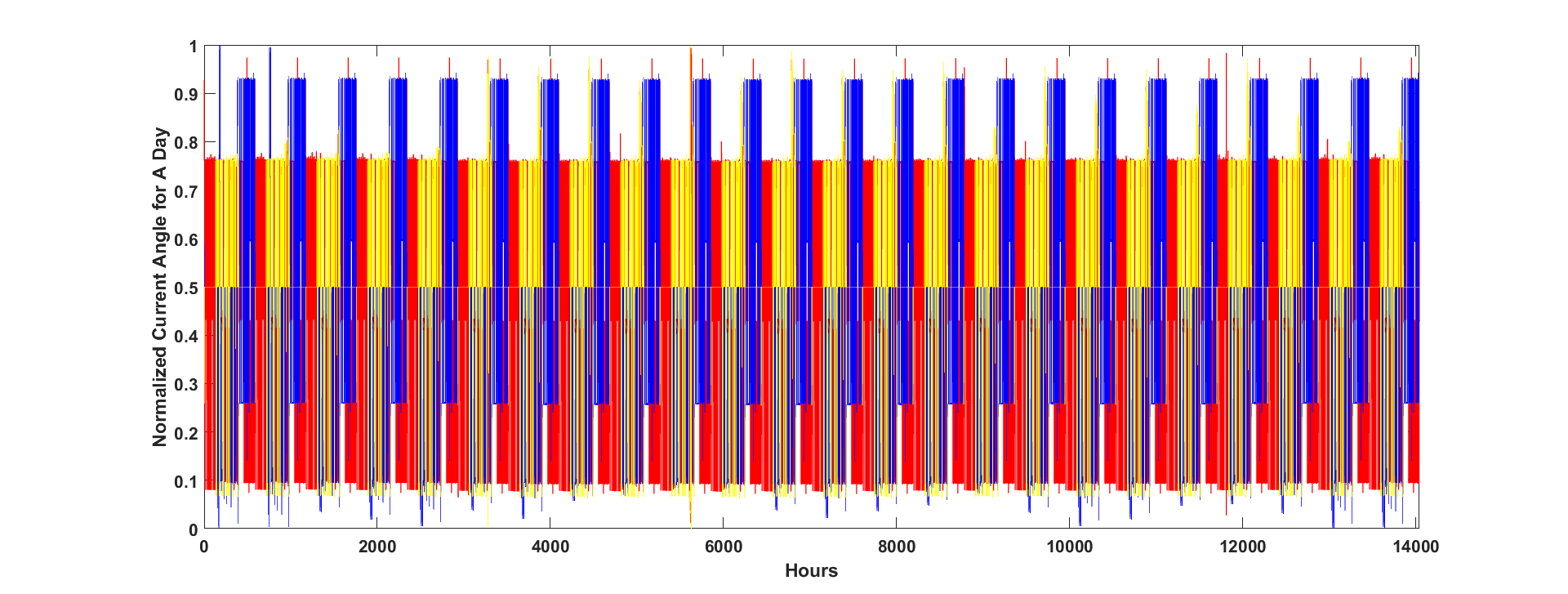}%
        \label{curr_ang}%
    }
    \caption{\small Simulation results of the DNN model for the AEP feeder for output voltage and current phasors.}
    \label{outputs}
    \vspace{-4mm}
\end{figure}

Fig \ref{outputs} represents the normalized output voltages, output currents, voltage angles, and current angles for a day for all three phases, respectively. As discussed in Section III, we have used the min-max normalization to process this data. {As can be seen in Table \ref{tab:my_label}, comparing the results of Fig \ref{outputs} with the expected values for voltage and current phasors provides very low error margins which shows the ability of the models to accurately predict the power flow solutions.}

Table \ref{tab:my_label} is the comparison of results for both networks. For IEEE 4-node, loads are modeled for three scenarios, i.e., constant PQ, constant impedance, and ZIP load. As shown in Table \ref{tab:my_label}, {the error margin is slightly higher for the MLP model when the load model is ZIP which shows that the complexity of the load poses a challenge for the model to capture the data patterns. Among the three models, the CNN model has better performance in terms of MSE which shows its ability to better capture the hidden patterns in the data. However,} different load models for the IEEE 4-node system do not {significantly} impact the performance of the proposed DNN models as the prediction errors in all three models are below 1\%. In the case of the AEP system, although the system is larger, the DNN models provide high accuracy making these models very precise at finding the PF solutions of larger systems. {Similar to the 4-bus system, the CNN model outperforms the other two models due to the ability of the convolutional layers to capture the relationships in the input data.} The very low values of MSE and MAE of the DNN models conclude that they could be applied to different systems with different sizes and load models making them a viable substitute for iterative models for finding load flow solutions. {The higher MAE values compared to MSE values of the models for both systems are due to the small errors throughout the predictions where MAE can capture these residuals more accurately than MSE.}
\begin{figure}[!h]
    \centering
    \includegraphics[clip, trim=0.0cm 0cm -7cm 0cm, width=0.6\textwidth]{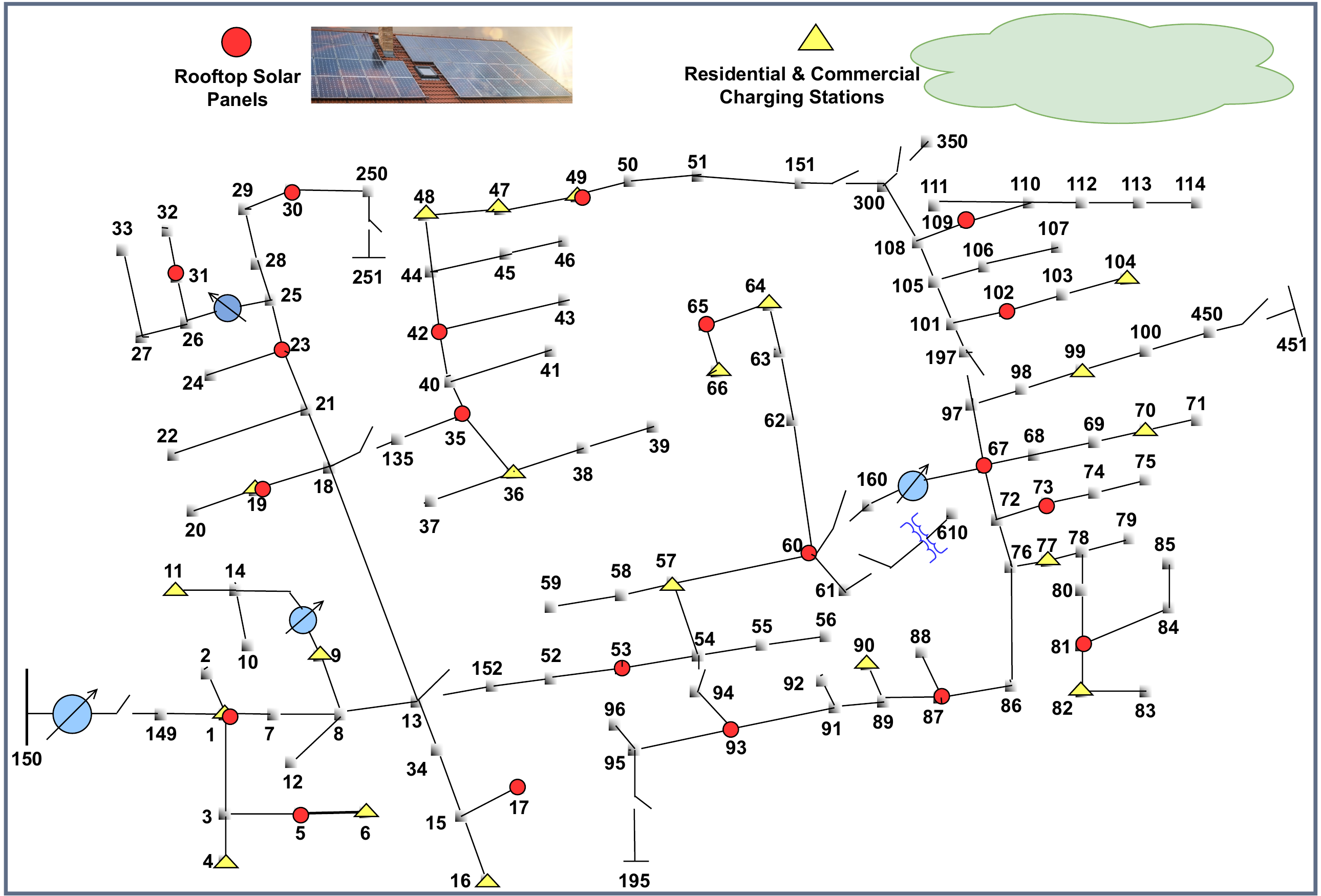}
    \caption{\small IEEE 123-Bus feeder with renewable energy sources}
   \label{fig: RES123}
\end{figure}
\subsection{Case 3: IEEE 123 bus}
To further test the feasibility of our proposed methodology, we use the IEEE 123-bus feeder shown in Fig. \ref{fig: RES123}. The data is generated using load-shapes of actual
load data of New York state and OpenDSS MATLAB COM platform. {The simulation settings for the IEEE 123-bus system in cases 3, 4, and 5 are the same as those of the AEP feeder in Case 2.}\par
It is evident from Table \ref{tab:my_label} that all three NNs resulted in very low MAE and MSE values in finding the PF solutions for the IEEE 123-bus system. {The higher MSE values compared to MAE values in the MLP and CNN models indicate that the prediction errors for a few data points are larger. Nevertheless, the magnitudes of these errors are not very large to significantly impact the overall performance of the models. On the other hand, the RBF net model has nearly similar performance to those of the 4-bus system and AEP feeder, showing very small consistent errors in the power flow predictions.} Predicting the load flow quantities with nearly zero error validates our claim that the proposed DNN models can predict the load flow quantities with nearly zero error in unbalanced distribution grids integrated with renewable energy sources and EV systems.

\subsection{Case 4: Topology Change}
We used the following methodologies to show the results of a topology change and measure the effectiveness of our proposed method:
\begin{itemize}
    \item Mixing the data: In this method, we trained and tested generated data
    before and after the topology change.  
    \item Generalization using Dropout: To overcome the risk of overfitting the dataset.
\end{itemize}
The results of the CNN model for topology change of an IEEE 123-bus system are shown in Table \ref{tab: Topology_PV_EV_results}. Comparing the results of Tables \ref{tab:my_label} and \ref{tab: Topology_PV_EV_results}, as expected, the MAE and MSE values increased with the change in the system topology. {When the system topology changes, the data distributions significantly differ from those in the original datasets without topology change. The higher MSE and MAE values demonstrate that the model has encountered new challenges in handling the new datasets producing the increased prediction errors.}
However, the error values are still low and confirm the validity of the method in cases of topology change.

\subsection{Case 5: PV and EV Integration }
\begin{figure}[!h]

    \includegraphics[width=0.5\textwidth]{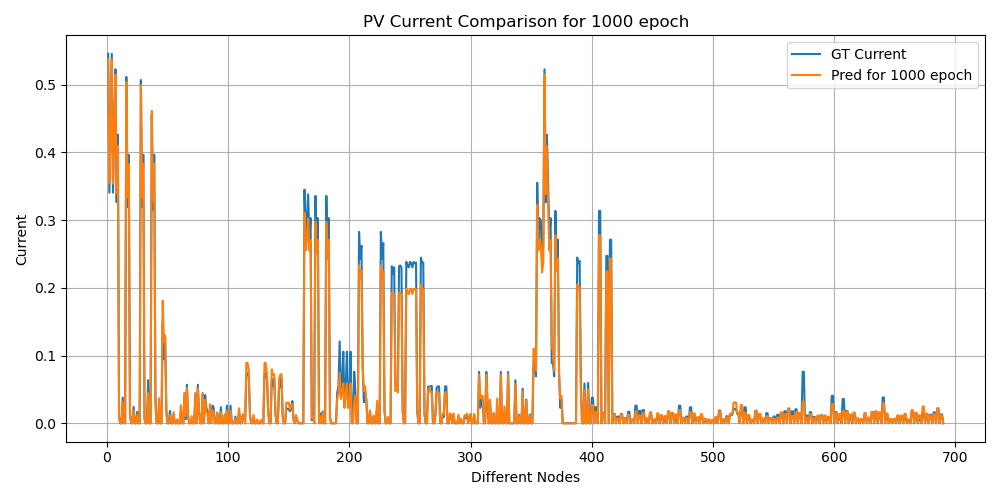}
     \caption{\small Currents for OpenDSS (GT) vs DNN }
   \label{fig: PV_GTvsAICurrents}
   \vspace{-3mm}
\end{figure}

In this case, we have predicted the power flow quantities in the IEEE 123-bus feeder which has multiple rooftop PVs (5-500 kW residential solar panels) and EVs using the deep neural network.
Results for PV and EV integration in an IEEE 123-bus system are shown in Table \ref{tab: Topology_PV_EV_results}. The predicted values of PV current and the ground truth values of OpenDSS when PVs are integrated into the system are shown in Fig. \ref{fig: PV_GTvsAICurrents}. {As can be seen in this figure, the CNN model has accurately predicted the PV currents with low error margins. In this scenario, due to the variability and randomness of the PV generations, the predicted values are slightly different during the periods of fluctuating currents.} 
As shown in Table \ref{tab: Topology_PV_EV_results}, with the integration of PV and EV into the system, the DNN model still provides highly accurate results compared to the ground truth values. {The higher prediction errors in cases of PV and EV integration arise from the uncertainty and variations in the input data as these sources are unpredictable and have random generation and consumption. These uncertainties in the data are reflected in the DNN model predictions with higher error margins.} {The overall low MSE and MAE values in cases of PV and EV integration} certify the applicability of the proposed DNN models in real cases with the presence of DERs and EV systems.

\begin{table}[!h]
  \caption{\small Prediction error} 
 \centering
 \scalebox{1.1}{
\begin{tabular}{|c|c|c|c|}
\hline
Test Cases                    & Scenarios       & \multicolumn{2}{c|}{CNN} \\ \hline
                              &                 & MAE         & MSE        \\ \hline
\multirow{3}{*}{IEEE 123 Bus} & Topology Change &  2.40$\%$       &  1.30$\%$      \\ \cline{2-4} 
                              & PV Integration  &  1.60$\%$       &  1.40$\%$       \\ \cline{2-4} 
                              & EV Integration  &  1.50$\%$       &  0.90 $\%$      \\ \hline
\end{tabular}}
\label{tab: Topology_PV_EV_results}
\end{table}

\section{Conclusion}

{In this paper, three DNN models, CNN, MLP, and RBFnet, are developed to solve the PF problem in unbalanced power distribution grids with mutual couplings. The proposed models can learn the nonlinear relationships in the PF equations and systems constraints to accurately approximate the solutions without the need for complex mathematical models or physics-based assumptions about the distribution systems. The training data was generated using the OpenDSS MATLAB COM interface. To make the DNN models applicable to real-world load scenarios, we considered various load models including constant PQ, constant impedance, and ZIP loads for the IEEE 4-bus system. In addition, the models are developed and tested for large-scale cases of AEP feeder and IEEE 123-bus system integrated with DERs and EVs to evaluate their implementations for large-scale real-world scenarios with the randomness and uncertainty of these resources. Testing the DNN models on different distribution networks with unbalanced configurations, mutual couplings, various load scenarios, integration of DERs and EV systems, and topology changes to find the PF solutions with low error margins demonstrated their robustness and showcased their potential to apply to the real-time operation of the distribution grids. Once trained, the DNN models can be employed as powerful tools to aid in the control and management of the distribution grids and help the system operators to have more informed insights into the behavior of the grids without relying on assumptions that might not be true under real-world cases.}
 
\bibliographystyle{IEEEtran}
\bibliography{access.bib}

\begin{thebibliography}{10}
\providecommand{\url}[1]{#1}
\csname url@samestyle\endcsname
\providecommand{\newblock}{\relax}
\providecommand{\bibinfo}[2]{#2}
\providecommand{\BIBentrySTDinterwordspacing}{\spaceskip=0pt\relax}
\providecommand{\BIBentryALTinterwordstretchfactor}{4}
\providecommand{\BIBentryALTinterwordspacing}{\spaceskip=\fontdimen2\font plus
\BIBentryALTinterwordstretchfactor\fontdimen3\font minus \fontdimen4\font\relax}
\providecommand{\BIBforeignlanguage}[2]{{%
\expandafter\ifx\csname l@#1\endcsname\relax
\typeout{** WARNING: IEEEtran.bst: No hyphenation pattern has been}%
\typeout{** loaded for the language `#1'. Using the pattern for}%
\typeout{** the default language instead.}%
\else
\language=\csname l@#1\endcsname
\fi
#2}}
\providecommand{\BIBdecl}{\relax}
\BIBdecl

\bibitem{pfapplication1}
J.~K. Skolfield and A.~R. Escobedo, ``Operations research in optimal power flow: A guide to recent and emerging methodologies and applications,'' \emph{European Journal of Operational Research}, vol. 300, no.~2, pp. 387--404, Jul 2022.

\bibitem{pfapplication2}
A.~Maheshwari, S.~Yog~Raj, and J.~Supriya, ``Flow direction algorithm-based optimal power flow analysis in the presence of stochastic renewable energy sources,'' \emph{Electric Power Systems Research}, vol. 216, p. 109087, Mar 2023.

\bibitem{mirzapour2022}
O.~Mirzapour and M.~Sahraei-Ardakani, ``Impacts of variable-impedance-based power flow control on renewable energy integration,'' \emph{arXiv preprint arXiv:2204.12642}, Apr 2022.

\bibitem{pfapplication3}
T.~Samakpong, O.~Weerakorn, and M.~M. Nimal, ``Optimal power flow incorporating renewable uncertainty related opportunity costs,'' \emph{Computational Intelligence}, vol.~38, no.~3, pp. 1057--1082, Jun 2022.

\bibitem{EV_effect}
A.~Soofi, R.~Bayani, and S.~Manshadi, ``Investigating the impact of electric vehicles on the voltage profile of distribution networks,'' in \emph{2022 IEEE Power \& Energy Society Innovative Smart Grid Technologies Conference (ISGT)}.\hskip 1em plus 0.5em minus 0.4em\relax IEEE, 2022, pp. 1--5.

\bibitem{radial_ring}
N.~Deyhimi, H.~Torkaman, M.~Shadaei, M.~Shabanirad, and M.~Kermani, ``Comparative multi-objective investigation of radial and ring distribution system in the presence of dgs,'' in \emph{2021 IEEE International Conference on Environment and Electrical Engineering and 2021 IEEE Industrial and Commercial Power Systems Europe (EEEIC/I\&CPS Europe)}.\hskip 1em plus 0.5em minus 0.4em\relax IEEE, 2021, pp. 1--6.

\bibitem{radial1}
F.~Zhang and S.~C. Carol, ``A modified newton method for radial distribution system power flow analysis,'' \emph{IEEE Transactions on Power Systems}, vol.~12, no.~1, pp. 389--397, Feb 1997.

\bibitem{radial2}
U.~Eminoglu and H.~M.~Hakan, ``A new power flow method for radial distribution systems including voltage dependent load models,'' \emph{Electric power systems research}, vol.~76, no. 1-3, pp. 106--114, Sep 2005.

\bibitem{radial3}
R.~D. Zimmerman and C.~Hsiao-Dong, ``Fast decoupled power flow for unbalanced radial distribution systems,'' \emph{IEEE Transactions on Power systems}, vol.~10, no.~4, pp. 2045--2052, Nov 1995.

\bibitem{PF_GNN_hansen}
J.~B. Hansen, S.~N. Anfinsen, and F.~M. Bianchi, ``Power flow balancing with decentralized graph neural networks,'' \emph{IEEE Transactions on Power Systems}, vol.~38, no.~3, pp. 2423--2433, 2023.

\bibitem{tradition1}
J.~S. Giraldo, P.~V. Pedro, C.~L. Juan, H.~N. Phuong, and G.~P. Nikolaos, ``A linear ac-opf formulation for unbalanced distribution networks,'' \emph{IEEE Transactions on Industry Applications}, vol.~57, no.~5, pp. 4462--4472, Jun 2021.

\bibitem{tradition2}
S.~Bolognani and Z.~Sandro, ``On the existence and linear approximation of the power flow solution in power distribution networks,'' \emph{IEEE Transactions on Power Systems}, vol.~31, no.~1, pp. 163--172, Feb 2015.

\bibitem{tradition3}
H.~Ahmadi, R.~M. José, and v.~M. Alexandra, ``A linear power flow formulation for three-phase distribution systems,'' \emph{IEEE Transactions on Power Systems}, vol.~31, no.~6, pp. 5012--5021, Feb 2016.

\bibitem{tradition4}
K.~Liu, W.~Chun, W.~Weizhang, C.~Yujie, and W.~Huicheng, ``Linear power flow calculation of distribution networks with distributed generation,'' \emph{IEEE Access}, vol.~7, pp. 44\,686--44\,695, Apr 2019.

\bibitem{tradition5}
A.~Bernstein and D.~Emiliano, ``Linear power-flow models in multiphase distribution networks,'' in \emph{2017 IEEE PES Innovative Smart Grid Technologies Conference Europe (ISGT-Europe)}.\hskip 1em plus 0.5em minus 0.4em\relax IEEE, 2017, pp. 1--6.

\bibitem{energy_LF}
K.~Amarasinghe, D.~Marino, and M.~Manic, ``Deep neural networks for energy load forecasting,'' in \emph{2017 IEEE 26th international symposium on industrial electronics (ISIE)}.\hskip 1em plus 0.5em minus 0.4em\relax IEEE, 2017, pp. 1483--1488.

\bibitem{short-term_LF}
A.~Haque and S.~Rahman, ``Short-term electrical load forecasting through heuristic configuration of regularized deep neural network,'' \emph{Applied Soft Computing}, vol. 122, p. 108877, Jun 2022.

\bibitem{two-sided}
M.~Jabbari~Zideh and S.~S. Mohtavipour, ``Two-sided tacit collusion: Another step towards the role of demand-side,'' \emph{Energies}, vol.~10, no.~12, p. 2045, Dec 2017.

\bibitem{iterative}
S.~S. Mohtavipour and M.~Jabbari~Zideh, ``An iterative method for detection of the collusive strategy in prisoner’s dilemma game of electricity market,'' \emph{EEJ Transactions on Electrical and Electronic Engineering}, vol.~14, no.~2, pp. 252--260, Feb 2019.

\bibitem{solar_battery}
R.~Nematirad, A.~Pahwa, B.~Natarajan, and H.~Wu, ``Optimal sizing of photovoltaic-battery system for peak demand reduction using statistical models,'' \emph{Frontiers in Energy Research}, vol.~11, p. 1297356, Dec 2023.

\bibitem{solar_radiation}
R.~Nematirad and A.~Pahwa, ``Solar radiation forecasting using artificial neural networks considering feature selection,'' in \emph{2022 IEEE Kansas Power and Energy Conference (KPEC)}.\hskip 1em plus 0.5em minus 0.4em\relax IEEE, 2022, pp. 1--4.

\bibitem{SE_cyber}
H.~Wang, J.~Ruan, G.~Wang, B.~Zhou, Y.~Liu, X.~Fu, and J.~Peng, ``Deep learning-based interval state estimation of ac smart grids against sparse cyber attacks,'' \emph{IEEE Transactions on Industrial Informatics}, vol.~14, no.~11, pp. 4766--4778, Feb 2018.

\bibitem{SE_application}
D.~Mukherjee, S.~Chakraborty, S.~Ghosh, and R.~Mishra, ``Application of deep learning for power system state forecasting,'' \emph{International Transactions on Electrical Energy Systems}, vol.~31, no.~9, p. e12901, Sep 2021.

\bibitem{PIConvAE}
M.~J. Zideh and S.~K. Solanki, ``Physics-informed convolutional autoencoder for cyber anomaly detection in power distribution grids,'' \emph{arXiv preprint arXiv:2312.04758}, Dec 2023.

\bibitem{PIML}
M.~J. Zideh, P.~Chatterjee, and A.~K. Srivastava, ``Physics-informed machine learning for data anomaly detection, classification, localization, and mitigation: A review, challenges, and path forward,'' \emph{IEEE Access (Early Access)}, Dec 2023.

\bibitem{security_mousavi}
A.~Mehrzad, M.~Darmiani, Y.~Mousavi, M.~Shafie-Khah, and M.~Aghamohammadi, ``A review on data-driven security assessment of power systems: Trends and applications of artificial intelligence,'' \emph{IEEE Access}, vol.~11, pp. 78\,671--78\,685, 2023.

\bibitem{pfmlsurvey}
H.~Fouad, A.~Kargarian, and A.~Mohammadi, ``A survey on applications of machine learning for optimal power flow,'' in \emph{2020 IEEE Texas Power and Energy Conference (TPEC)}.\hskip 1em plus 0.5em minus 0.4em\relax IEEE, 2020, pp. 1--6.

\bibitem{pfphysics}
X.~Lei, Y.~Zhifang, Y.~Juan, Z.~Junbo, G.~Qian, and Y.~Hongxin, ``Data-driven optimal power flow: A physics-informed machine learning approach,'' \emph{IEEE Transactions on Power Systems}, vol.~36, no.~1, pp. 346--354, Jun 2020.

\bibitem{pfmlarx}
N.~Guha, W.~Zhecheng, W.~Matt, and M.~Arun, ``Machine learning for ac optimal power flow,'' \emph{arXiv preprint arXiv:1910.08842}, 2019.

\bibitem{DeepOPF}
X.~Pan, C.~Minghua, Z.~Tianyu, and H.~Steven, ``Deepopf: A feasibility-optimized deep neural network approach for ac optimal power flow problems,'' \emph{IEEE Systems Journal}, Sep 2022.

\bibitem{pfDeepRL}
Y.~Zhou, Z.~Bei, X.~Chunlei, L.~Tu, D.~Ruisheng, S.~Di, W.~Zhiwei, and L.~Wei-Jen, ``A data-driven method for fast ac optimal power flow solutions via deep reinforcement learning,'' \emph{Journal of Modern Power Systems and Clean Energy}, vol.~8, no.~6, pp. 1128--1139, Nov 2020.

\bibitem{deepak_thesis}
D.~Tiwari, ``Application of artificial intelligence in three phase unbalanced smart power distribution grid,'' \emph{Graduate Theses, Dissertations, and Problem Reports. 8005}, 2021.

\bibitem{multi-MG}
H.~Xiao, W.~Pei, L.~Wu, L.~Ma, T.~Ma, and W.~Hua, ``A novel deep learning based probabilistic power flow method for multi-microgrids distribution system with incomplete network information,'' \emph{Applied Energy}, vol. 335, p. 120716, Apr 2023.

\bibitem{lifting_LR}
L.~Guo, Y.~Zhang, X.~Li, Z.~Wang, Y.~Liu, L.~Bai, and C.~Wang, ``Data-driven power flow calculation method: A lifting dimension linear regression approach,'' \emph{IEEE Transactions on Power Systems}, vol.~37, no.~3, pp. 1798--1808, Sep 2021.

\bibitem{GNN_realistic}
L.~Böttcher, H.~Wolf, B.~Jung, P.~Lutat, M.~Trageser, O.~Pohl, A.~Ulbig, and M.~Grohe, ``Solving ac power flow with graph neural networks under realistic constraints,'' \emph{arXiv preprint arXiv:2204.07000}, 2022.

\bibitem{support_matrix}
J.~Yuan and Y.~Weng, ``Support matrix regression for learning power flow in distribution grid with unobservability,'' \emph{IEEE Transactions on Power Systems}, vol.~37, no.~2, pp. 1151--1161, Aug 2021.

\bibitem{Nonlinearity_Adaptive}
Y.~Zhang, L.~Guo, Y.~Liu, Z.~Wang, X.~Li, L.~Bai, and C.~Wang, ``Nonlinearity-adaptive data-driven power flow constraint for distribution network optimization,'' \emph{IEEE Transactions on Power Systems}, Jan 2023, doi: 10.1109/TPWRS.2023.3236142.

\bibitem{Data-driven_RBM}
P.~Algikar, Y.~Xu, S.~Yarahmadi, and L.~Mili, ``A robust data-driven process modeling applied to time-series stochastic power flow,'' \emph{IEEE Transactions on Power Systems}, Jan 2023, doi: 10.1109/TPWRS.2023.3238385.

\bibitem{robust_mapping}
J.~Yu, Y.~Weng, and R.~Rajagopal, ``Robust mapping rule estimation for power flow analysis in distribution grids,'' in \emph{017 North American Power Symposium (NAPS)}.\hskip 1em plus 0.5em minus 0.4em\relax IEEE, 2017, pp. 1--6.

\bibitem{OPF_gao_GNN}
M.~Gao, J.~Yu, Z.~Yang, and J.~Zhao, ``A physics-guided graph convolution neural network for optimal power flow,'' \emph{IEEE Transactions on Power Systems}, vol.~39, no.~1, pp. 380--390, 2024.

\bibitem{PI_opf_nellikkath}
R.~Nellikkath and S.~Chatzivasileiadis, ``Physics-informed neural networks for ac optimal power flow,'' \emph{Electric Power Systems Research}, vol. 212, p. 108412, 2022.

\bibitem{decentralized_ADMM}
T.~W. Mak, M.~Chatzos, M.~Tanneau, and P.~Van~Hentenryck, ``Learning regionally decentralized ac optimal power flows with admm,'' \emph{IEEE Transactions on Smart Grid}, vol.~14, no.~6, pp. 4863--4876, 2023.

\bibitem{PF_UC_alyssa}
A.~Kody, S.~Chevalier, S.~Chatzivasileiadis, and D.~Molzahn, ``Modeling the ac power flow equations with optimally compact neural networks: Application to unit commitment,'' \emph{Electric Power Systems Research}, vol. 213, p. 108282, 2022.

\bibitem{typed_GNN}
T.~B. Lopez-Garcia and J.~A. Domínguez-Navarro, ``Power flow analysis via typed graph neural networks,'' \emph{Engineering Applications of Artificial Intelligence}, vol. 117, p. 105567, 2023.

\bibitem{bayesian_DCMG}
P.~R. Jeyaraj, S.~P. Asokan, and A.~C. Karthiresan, ``Physics-informed neural networks for ac optimal power flow,'' \emph{Electric Power Systems Research}, vol. 205, p. 107730, 2022.

\bibitem{thukaram1999robust}
D.~Thukaram, H.~W. Banda, and J.~Jerome, ``A robust three phase power flow algorithm for radial distribution systems,'' \emph{Electric Power Systems Research}, vol.~50, no.~3, pp. 227--236, 1999.

\bibitem{FBS_solanki}
S.~Khushalani, J.~M. Solanki, and N.~N. Schulz, ``Optimized restoration of unbalanced distribution systems,'' \emph{IEEE Transactions on Power Systems}, vol.~22, no.~2, pp. 624--630, Apr 2007.

\bibitem{kersting2017distribution}
W.~H. Kersting, \emph{Distribution system modeling and analysis}.\hskip 1em plus 0.5em minus 0.4em\relax CRC press, 2017.

\bibitem{zimmerman1995fast}
R.~D. Zimmerman and H.-D. Chiang, ``Fast decoupled power flow for unbalanced radial distribution systems,'' \emph{IEEE Transactions on Power systems}, vol.~10, no.~4, pp. 2045--2052, 1995.

\bibitem{teng2003direct}
J.-H. Teng, ``A direct approach for distribution system load flow solutions,'' \emph{IEEE Transactions on power delivery}, vol.~18, no.~3, pp. 882--887, 2003.

\bibitem{dugan2016reference}
R.~C. Dugan, ``Reference guide,'' \emph{The Open Distribution System Simulator (OpenDSS). EPRI}, 2016.

\bibitem{linear_PF}
Y.~Wang, N.~Zhang, H.~Li, J.~Yang, and C.~Kang, ``Linear three-phase power flow for unbalanced active distribution networks with pv nodes,'' \emph{CSEE Journal of Power and Energy Systems}, vol.~3, no.~3, pp. 321--324, 2017.

\bibitem{CNN_introduction}
J.~Wu, ``Introduction to convolutional neural networks,'' \emph{National Key Lab for Novel Software Technology. Nanjing University. China}, vol.~5, no.~23, p. 495, May 2017.

\bibitem{duda2012pattern}
R.~O. Duda, P.~E. Hart, and D.~G. Stork, \emph{Pattern classification}.\hskip 1em plus 0.5em minus 0.4em\relax John Wiley \& Sons, 2012.

\end{thebibliography}

\end{document}